\documentclass[12pt]{spieman}
\usepackage{amsmath,amsfonts,amssymb}
\usepackage{graphicx}
\usepackage{setspace}
\usepackage{tocloft}

\title{Whispering-gallery-mode barcode-based broadband sub-femtometer-resolution spectroscopy with an electro-optic frequency comb}

\author[1,2,$\dagger$]{Bingxin Xu}
\author[1,$\dagger$]{Yangyang Wan}
\author[1,*]{Xinyu Fan}
\author[1]{Zuyuan He}

\affil[1]{State Key Laboratory of Advanced Optical Communication Systems and Networks, Department of Electronic Engineering, Shanghai Jiao Tong University, Shanghai 200240, China}
\affil[2]{Current address: Max-Planck Institute of Quantum Optics, Hans-Kopfermann-Stra{\ss}e 1, 85748, Garching, Germany}

\affil[$\dagger$]{These authors contributed equally to this work.}

\cftpagenumbersoff{figure}
\cftpagenumbersoff{table} 
\begin{document} 
\maketitle

\begin{abstract}
Spectroscopy is the basic tool for studying molecular physics and realizing bio-chemical sensing. 
However, it is challenging to realize sub-femtometer resolution spectroscopy over broad bandwidth.
In this paper, broadband and high-resolution spectroscopy with calibrated optical frequency is demonstrated by bridging the fields of speckle patterns and electro-optic frequency comb (EOFC). 
A novel wavemeter based on whispering-gallery-mode (WGM) speckles (or WGM barcodes) is proposed to link the frequency of a tunable continuous-wave (CW) laser to an optical reference provided by an ultra-stable laser.
The ultra-fine comb lines generated from the CW laser sample the spectrum with sub-femtometer resolution. Measurement bandwidth is far extended by 
performing sequential acquisitions, since the centre optical frequency of EOFC is absolutely determined by WGM speckle-based wavemter. This approach fully utilizes the advantages of two fields to realize 0.8-fm resolution with a fiber laser and 80-nm bandwidth with an external cavity diode laser. The spectroscopic measurements of an ultrahigh-Q cavity and the HCN gas absorption is demonstrated, which shows the potentials of this compact system with high resolution and broad bandwidth for more applications.
\end{abstract}


{\noindent \footnotesize\textbf{*}Xinyu Fan,  \linkable{fan.xinyu@sjtu.edu.cn} }

\begin{spacing}{1}   

\section{Introduction}
Spectroscopy is the essential tool to investigate the structures of atoms and molecules, which has outstandingly contributed to atomic and molecular physics, analytical chemistry, and molecular biology. It also gained enormous significance in the fields of optical sensing, environmental study, and medical diagnostic. Spectroscopic measurements with high resolution, broad bandwidth and calibrated optical frequency are fundamentally important for high-precision investigations of multiple samples.

Spectrometers, widely implemented in spectroscopic measurement, are facilitated by using speckle patterns~\cite{2022cao,2021wan2}. The speckle pattern generated in disordered media is distinct at each wavelength to reconstruct the input spectrum. Random interference or scattering generate speckle patterns in space during the propagation of light in multimode fiber~\cite{2012redding,2013redding,2014redding,2016coluccelli,2019bruce,2020gupta,2020wang}, integrating sphere~\cite{2017metzger}, or some other materials~\cite{2016redding,2020kwak,2020yi,2021zhang}. Similarly, speckle patterns in time domain are generated from Rayleigh scattered lightwave in single-mode fiber, which can be detected by using a single photodetector (PD) instead of using camera or PD array~\cite{2020zhang,2021Wan}.
The compact speckle-based spectrometers have reached picometer resolution outperforming the state-of-art grating spectrometers~\cite{2014redding,2016coluccelli}. It is challenging to further reduce the spectral resolution. 
The determination of wavelength with higher precision may be realized by using speckle-based wavemeter. Sub-femtometer resolutions have been demonstrated respectively by using cross-correlation algorithm~\cite{2021Wan}, principal component analysis~\cite{2019bruce}, convolutional neutral network~\cite{2020gupta}. The speckle-based wavemeter only determine sparse discrete wavelength rather than an optical spectrum. Besides, it is not suitable to be used in swept laser spectroscopy, since the speckle patterns may be distorted. Therefore, the current implementations of speckle-based reconstructive schemes are incapable to realize high-resolution spectroscopy.

Optical frequency comb (OFC)~\cite{2002Udem,2010Diddams} is invented from femtosecond fiber laser for precision metrology by building a link between optical and radio frequency. Although laser spectroscopy is promoted from the assistance of OFC~\cite{2009del,giorgetta2010,Yang2019}, the equidistant coherent lines from OFC may be more powerful tools for simultaneously measuring broadband spectra as direct comb spectroscopy. Several techniques have been proposed for resolving the comb lines in broad bandwidth, including using a virtually imaged phase array~\cite{diddams2007}, using a scanning Fabry-Perot (F-P) cavity~\cite{gohle2007frequency}, Fourier transform spectrometer~\cite{mandon2009}, dual-comb spectrometer~\cite{2016Coddington}, and speckle-based fiber spectrometer~\cite{2016coluccelli}. In the last decade, more comb sources have been demonstrated with different repetition frequencies at more wavelength regions~\cite{2007DelHaye,2012hugi}. 
Electro-optic frequency combs (EOFCs)~\cite{2015bao,2016long,2016millot,2018carlson,XuTwoMod2019,XuInter2022} are attractive tools in high-resolution spectroscopy for providing ultra-fine comb lines. 
EOFC is generated by electro-optic modulation of a continuous-wave laser, which provides the agility of both centre frequency and repetition frequency.
The repetition rate of EOFC, practically spectral resolution in many direct comb spectroscopy, is not limited by physical cavity length.
Spectroscopy with sub-MHz resolution is facilitated by digitally-generated ultra-fine EOFCs~\cite{2015bao,2018yan,2019long,WangFast2019,XuInLock2021}, which reaches high refresh rate using coherent detection.
Relatively, the bandwidth of EOFC is limited by the electrical driving synthesizer. Although the measurement bandwidth of EOFC may be broadened 
from the frequency agility provided by the tunable seed laser, the frequency precision is far behind the requirement for sub-femtometer resolution spectroscopy.

Here, our work leverages the advantages of speckle-based wavemeter and ultra-fine EOFC to achieve ultra-high resolution spectroscopy with calibrated absolute frequency over a broad bandwidth.
A novel wavelength (or optical frequency) determination technique based on whispering-gallery-mode (WGM) barcodes reaching sub-femtometer resolution is proposed.
WGM barcodes (or WGM speckle patterns) are composed of complex WGM resonances with different depths, spacings, linewidths, and resonant wavelengths, which  are firstly proposed for temperature sensing~\cite{2021liao}.  
A continuous-wave (CW) laser is used to obtain measurement speckle for wavelength determination and generate ultra-fine EOFC for high-resolution spectroscopy. Compared with a reference speckle calibrated by an ultra-stable laser, the measurement speckle determine the absolute wavelength of the CW laser thanks to the ability of two-tone demodulation~\cite{2021wan3}. High-resolution spectrum of sample and accurately determined frequency of each comb lines are simultaneously achieved in millisecond measurement time. 
Broadband measurement is realized by doing sequential acquisitions at different centre wavelength without the requirement of precise wavelength adjustment of the tunable CW laser.
As an application of the scheme, spectroscopic measurement of a high-Q fiber F-P cavity with 0.8 fm resolution in 1 nm bandwidth and HCN gas cell with 8 fm resolution in 60 nm bandwidth are respectively demonstrated with a fiber laser and an external cavity diode laser (ECDL).
The proposed method bridges the fields of speckle-based wavemeter and comb-based spectroscopy, which shows promising prospects for precise spectroscopy with broad bandwidth.

\section{Principle and Methods}

The principle of speckle-based spectroscopy with an EOFC is shown in Fig. 1. 
The WGM speckle patterns (or WGM barcodes) are composed of complex WGMs excited in a silica micro-rod resonator. A tapered fiber is used to couple the light into and out of the WGM resonator. The WGM speckle patterns are unique at each wavelength (or optical frequency), since the depths, spacings, linewidths, and resonant wavelengths of WGM resonances are different. Hence, wavelength can be determined from such wavelength-dependent speckles. Cross-correlation algorithm is used to find the relative location of the speckle compared with a broadband reference speckle. 
The reference and measurement speckles are obtained by swept laser configuration with the setup shown in Fig. 2. The reference WGM speckle is recorded before measurement. A linearly frequency swept RF signal with a sweep range of 10 GHz is applied on the single-sideband modulator (SSBM). The CW laser is modulated and passes through the WGM resonator. A PD with 120 MHz bandwidth and an OSC with 250 MSa/s sampling rate are used for detection. The WGM resonances in 10 GHz banwidth is measured, which is regarded as an individual WGM speckle. The power fluctuation during the frequency sweeping is eliminated by introducing a channel without the resonator. Several WGM speckles over a broad bandwidth are obtained by adjusting the centre frequency of the tunable CW laser with a step of about 5 GHz or 40 pm. The spectra of adjacent acquisitions have an overlapping region, which is used for the stitching of these acquisitions to get a reference speckle covering the whole bandwidth. The stitching points are calculated by using cross-correlation algorithm.
The reference speckle can be regarded as a ``ruler'' in optical frequency domain by using the speckle of an ultra-stable laser to calibrate the absolute frequency. Hence, the absolute frequency of the CW laser is obtained from relative location. 

\begin{figure*}[tb]
\centering
\includegraphics[width=15cm]{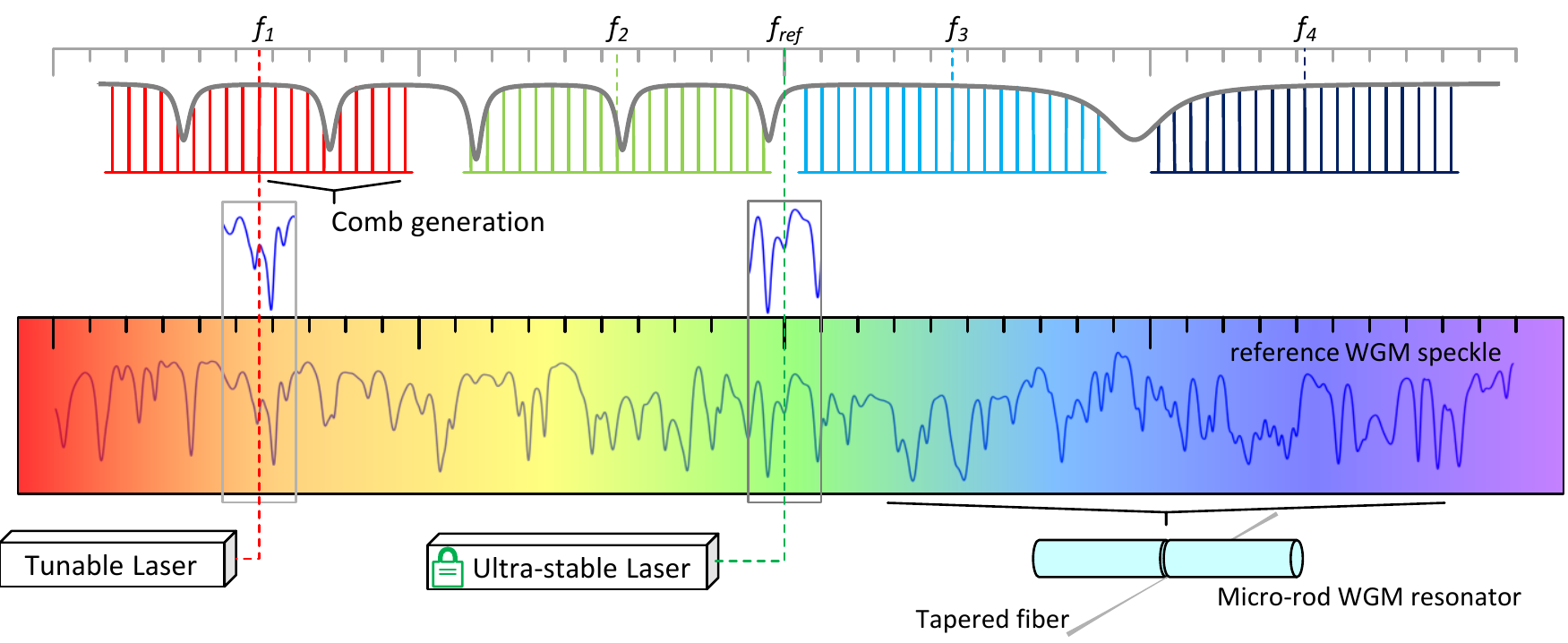}
\caption{\textbf{Speckle-based spectroscopy with an electro-optic frequency comb.} A mass of whispering-gallery modes (WGMs) is excited from a micro-rod resonator attached with a tapered fiber. The WGM speckle (or WGM barcode) is the WGM resonances around one wavelength (or optical frequency) for wavelength determination. The WGM speckle in whole bandwidth is a broadband ``ruler'' with the calibration of the absolute frequency with a reference laser. A tunable continuous-wave laser, whose frequency is determined from the speckle on the ``ruler'', is used to generate an ultra-fine comb. High-resolution spectroscopy is performed with an ultra-fine electro-optic frequency comb. The bandwidth is extended to cover whole range by tuning the centre frequency.}
\label{fig1}
\end{figure*}
\begin{figure*}[tb]
\centering
\includegraphics[width=10cm]{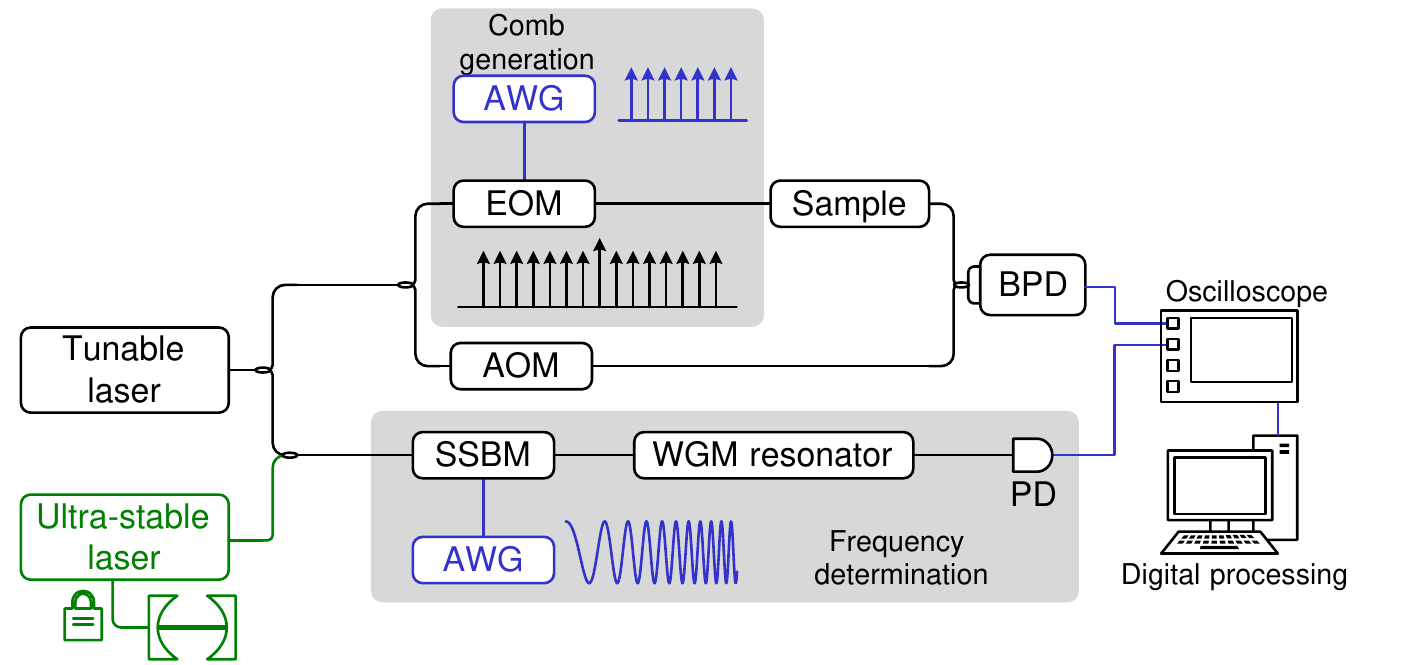}
\caption{\textbf{The experimental setup of the speckle-based spectroscopy with an electro-optic frequency comb.} AWG: arbitrary waveform generator; EOM: electro-optic modulator; SSBM: single-sideband modulator;WGM: whispering-gallery-mode; PD: photo-detector; BPD: balanced photo-detector. }
\label{fig2}
\end{figure*}

The EOFC generated from the CW laser records the spectrum of the sample, and its comb lines are resolved in an interferometer. Extending the bandwidth of high-resolution spectroscopy is enabled due to the calibrated frequency and the tunability of the CW laser. By adjusting the centre frequency of the tunable laser, the broadband precise spectra are acquired from EOFC centred at different frequencies. Importantly, precise wavelength tuning is not required for high-resolution spectroscopy, since the absolute wavelength is determined from the speckle patterns. The bandwidth of spectroscopy is no longer limited by the bandwidth of EOFC but by the tunable range of the CW laser considering much broader range of the WGM speckle. 

The experimental setup is shown in Fig. 2. The WGM resonator is a silica micro-rod with a diameter of 1.5 mm fabricated by using CO$_2$ laser. The Q-factor of the resonator reaches over 10$^7$. Higher Q-factor contributes to increasing the resolution of frequency determination by both reducing the FWHM of the cross-correlation peak and the noise floor. The quantified simulation results are detailed in Supplementary 1, Section 2. The average insertion loss of the resonator is about 10.2 dB.

The tunable laser and the ultra-stable laser are coupled to obtain their WGM speckles. A SSBM is driven by a linearly-swept radio frequency (RF) signal. The dual-wavelength lightwaves are frequency-modulated to record several WGM resonances of the WGM resonator around their centre wavelengths. The WGM speckles are detected and digitized by using a PD and an oscilloscope (OSC). The data, represented as the superposition of two WGM speckles, is compared with the reference speckle by using cross-correlation algorithm to determine the absolute frequency of the tunable laser. Another partial output of tunable laser is modulated by using an electro-optic modulator (EOM) to generate EOFC. The driven RF signal generated from an arbitrary waveform generator (AWG) is a periodic waveform with a flat spectrum, which can be regarded as an ``electrical comb''. An EOFC with twice of the RF bandwidth is generated. The repetition frequency set by the period of the driven signal reaches sub-MHz level for high-resolution spectroscopy. The comb lines recording the spectrum of the sample are measured by using an interferometer. An acousto-optic modulator (AOM) introduces a centre frequency shift to distinguish the positive and negative sidebands of the EOFC. The interferograms are detected by using a balanced photodetector (BPD). The comb lines are resolved after Fourier transformation in digital processing. The acquisition of WGM speckle and comb spectrum is realized within 1 ms, regarded as one acquisition at one centre frequency. Longer measurement time contributes to improving the signal-to-noise ratio (SNR). A sequence of acquisitions by changing the centre frequency of the tunable laser obtains a broad bandwidth. In following experiments, a fiber laser with low frequency noise, and an ECDL with broad tuning range, respectively serve as the tunable laser to reach 100 kHz spectral resolution and 80 nm bandwidth.

\section{Results}
\subsection{High resolution of the system}

\begin{figure*}[tb]
\centering
\includegraphics[width=15cm]{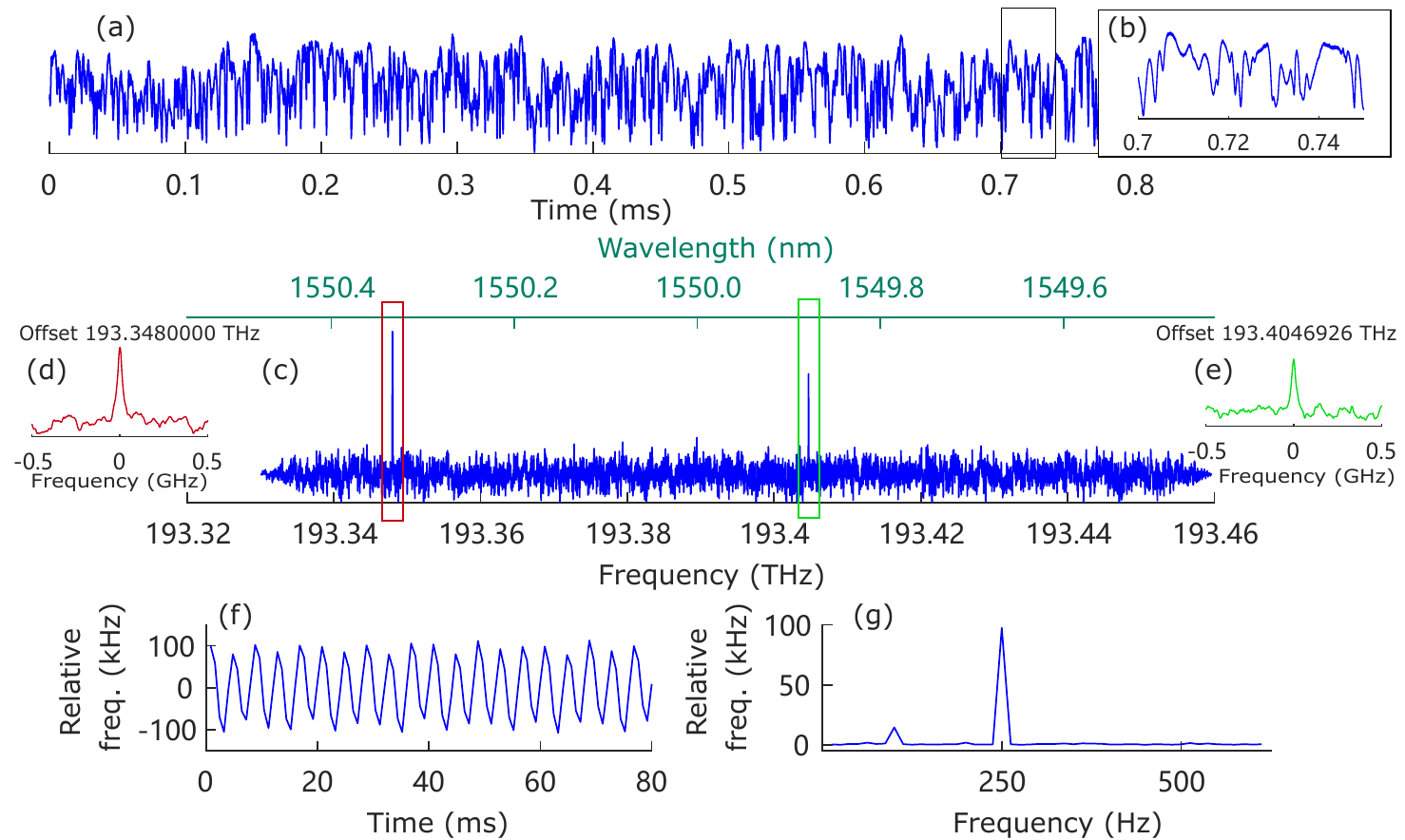}
\caption{\textbf{Frequency determination for the fiber laser with a resolution of 100 kHz in 120 GHz bandwidth.} (a) The speckle recorded in 0.8 ms, corresponding to a swept frequency range of 8 GHz. (b) A zoom-in figure of (a) in 0.05 ms. (c) Cross-correlation result between the measurement speckle and the reference speckle over a bandwidth of 120 GHz. (d) and (e) are zoom-in correlation peaks of the reference laser and the tunable laser. The frequency of the tunable laser is determined to be 193.4046926 THz. The full widths at half maximum (FWHMs) of two peaks are both about 41 MHz. (f) Frequency readouts of the tunable laser with frequency modulation. (g) The Fourier transformation reveals the amplitude and frequency of the modulation are respectively 100 kHz and 250 Hz. The sampling rate of the frequency readout is 1.25 kHz.}
\label{fig3}
\end{figure*}

A fiber laser serves as the tunable laser in the demonstration of high-resolution measurement. The superposition of the WGM speckles of the tunable laser and the ultra-stable laser in 0.8 ms is shown in Fig. 3(a). The range in optical domain is 8 GHz, which contains totally more than one hundred WGM resonances. These resonances have different depths and widths, as shown in the zoom-in image (Fig. 3(b)) in 50 $\mu$s. The WGM speckles composed of a mass of WGM resonances can be regarded an approximately stochastic curve at certain wavelength region. Cross-correlation results comparing this measurement speckle and the reference speckle are shown in Fig. 3(c). The bandwidth is limited by the tunable range of the fiber laser to be 120 GHz. The left peak, shown in Fig. 3(d) in red, represents the location of the ultra-stable laser at 193.3480000 THz, which calibrates the frequency of the whole cross-correlation results. The frequency of the fiber laser is determined to be 193.4046926 THz with a resolution of 100 kHz according to the right peak of the correlation-curve as shown in Fig. 3(e). The peak location determination is based on Lorentz fitting. The full widths at half maximum (FWHMs) of two peaks are both about 41 MHz.

The resolution of frequency determination is evaluated by measuring a frequency-modulated fiber laser. An AOM is introduced to modulate the centre frequency. The readouts of the relative frequency obtained from the WGM speckles are shown in Fig. 3(f). The sinusoidal frequency modulation with an amplitude of 100 kHz is well demodulated. The refresh rate of the frequency determination is 1.25 kHz. The Fourier transformation shown in Fig. 3(g) reveals the modulation frequency at 250 Hz with a SNR of 261.

\subsection{Spectroscopic measurement of a high-Q Fabry-Perot cavity}

\begin{figure*}[tb]
\centering
\includegraphics[width=15cm]{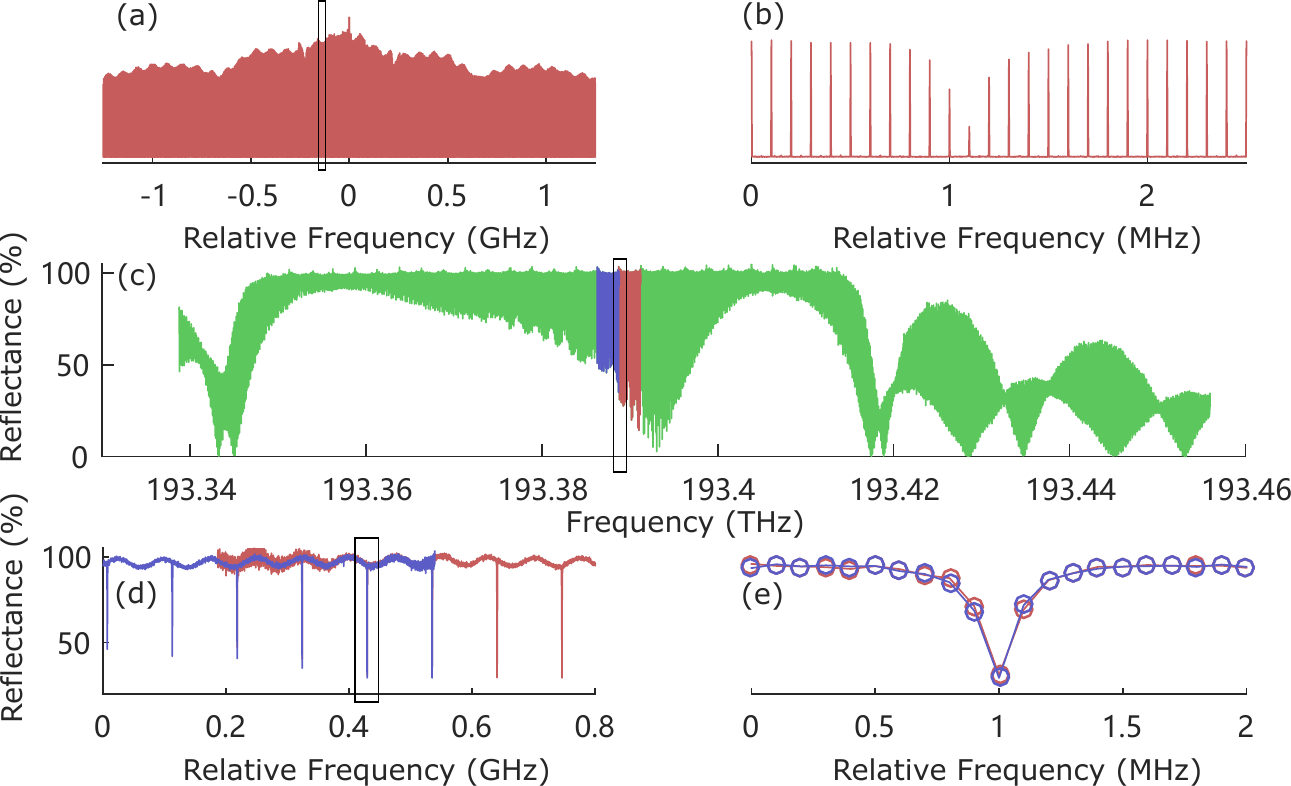}
\caption{\textbf{Spectroscopic measurement of the reflection spectrum of a fiber Fabry-Perot cavity.} (a) Resolved spectrum of the EOFC with a bandwidth of 2.5 GHz by Fourier transformation of data in 1 ms recording time. (b) A zoom-in figure of (a) with resolved comb lines and one recorded resonance. The line-spacing, corresponding to the spectral resolution, is 100 kHz. (c) The reflectance spectrum of a fiber Fabry-Perot cavity in 117 GHz bandwidth. The spectrum is composed of 47 acquisitions by changing the centre frequency of the tunable laser. Each acquisition contains 25000 resolved comb lines with 100 kHz resolution over 2.5 GHz bandwidth. Two acquisitions with adjacent centre frequencies are drawn in red and blue. (d) A zoom-in figure of (c) for an overlapped region in 0.8 GHz. The free spectral range of the fiber cavity is measured to be 105.6 MHz. (e) A zoom-in figure of a fiber cavity resonance. The line-width of the resonance is 250 kHz, corresponding to a Q-factor of 7.7$\times$10$^8$. The signal-to-noise ratio is 227. }
\label{fig4}
\end{figure*}

The EOFC generated from the fiber laser by electro-optic modulation is measured in the heterodyne interferometer. The interferograms in 1 ms recording time are Fourier transformed to obtain the spectrum of EOFC as shown in Fig. 4(a). 25000 comb lines covering a bandwidth of 2.5 GHz are resolved. The centre frequency in RF domain is 80.025 MHz corresponding to the frequency shift introduced by AOM, and the centre frequency in optical domain is determined by its WGM speckle. A zoom-in figure in 2.5 MHz bandwidth is shown in Fig. 4(b), in which the linewidth of comb line is Fourier-transform-limited to be 1 kHz. The line-spacing is 100 kHz, corresponding to the resolution of spectroscopy. The sample is a fiber F-P cavity composed of two fiber Bragg gratings. Its reflectance spectrum is measured by the EOFC. One resonance of the F-P cavity is also represented in Fig. 4(b).

The spectroscopy with extended bandwidth is performed by adjusting the centre frequency of the fiber laser. The whole reflectance spectra in 117 GHz bandwidth are obtained from a sequence of 47 acquisitions, as shown in Fig. 4(c). Each acquisition resolves 25000 comb lines in 1 ms recording time, and the absolute frequency is determined in 0.8 ms. Totally more than 10$^6$ spectral points are demodulated. The cavity resonances are in the reflective band around 193.39 THz in 0.5 THz bandwidth. Two adjacent acquisitions (respectively in blue and red) containing several narrow resonances are shown in Fig. 4(d). These resonances in overlapping region are well matched thanks to the absolute frequency calibration. The SNR is 227, calculated from the standard deviation of the baseline. The ripple on the baseline introduced from the fiber F-P cavity is eliminated by using a fitting process in calculation. The free spectral range of the fiber cavity is measured to be 105.6 MHz. One of the resonance with a linewidth of 250 kHz is shown in Fig. 4(e), corresponding to a cavity Q-factor of 7.7$\times$10$^8$. The ultrahigh-Q cavities, as essential tools in precise optical sensing and nonlinear optics, can be properly characterized in the proposed technique.

\subsection{Broad bandwidth of the system}
\begin{figure*}[tb]
\centering
\includegraphics[width=15cm]{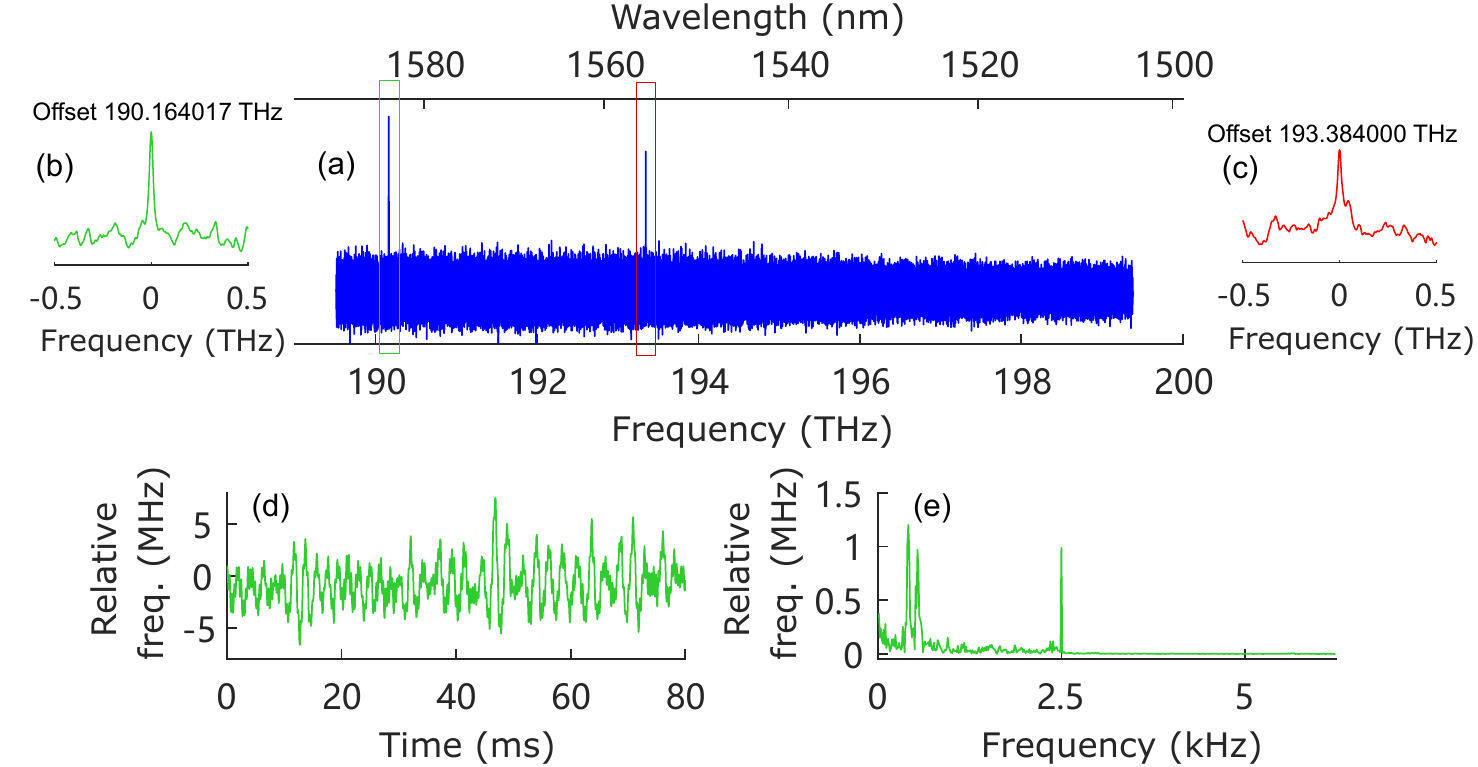}
\caption{\textbf{Frequency determination for the external cavity laser diode with a resolution of 1 MHz in 9.5 THz bandwidth.} (a) Cross-correlation result between the measurement speckle and reference speckle over a bandwidth of 9.5 THz. (b) A zoom-in correlation peak of the ECDL. The frequency is determined to be 190.164017 THz. (c) A zoom-in correlation peak of the reference laser. The FWHMs of two peaks are respectively 31 MHz and 54 MHz. (d) Frequency readouts of the ECDL with frequency modulation. (e) The Fourier transformation of (d). The amplitude and frequency of the modulation are respectively 1 MHz and 2.5 kHz. The low frequency region is the frequency noise of the ECDL. The averaged frequency error of ECDL in 1 ms (recording time for following spectroscopy experiment) is 0.48 MHz. The sampling rate of frequency determination is 12.5 kHz. }
\label{fig5}
\end{figure*}

The broad bandwidth of the system is performed by using an ECDL as the tunable laser. The tunable range of the laser is extended to 9.5 THz (or 76 nm). Similarly, the cross-correlation result between the measurement speckle and reference speckle is shown in Fig. 5(a). The green curve shown in Fig. 5(b) and the red curve shown in Fig. 5(c) respectively represent the centre frequencies of the ECDL and the ultra-stable laser. The frequency of the ECDL is determined to be 190.164017 THz. The resolution is 1 MHz, limited by the frequency noise of the ECDL rather than instrumental performance. 

To evaluate the resolution, frequency modulation with an amplitude of 1 MHz is also introduced by using AOM. The frequency readouts are shown in Fig. 5(d). The refresh rate of the frequency determination is increased to 12.5 kHz. The Fourier transformation reveals the frequency modulation with an amplitude of 1 MHz and a frequency of 2.5 kHz, as shown in Fig. 5(e). Intrinsic frequency noises of the ECDL in low frequency region with about 1-MHz amplitudes are also observed. Therefore, the readouts in Fig. 5(d) are the superposition of frequency modulation and frequency noise. The frequency noise is about 0.48 MHz in 1 ms, which is the limitation of spectral resolution for further spectroscopic measurement.

\subsection{Spectroscopic measurement of HCN gas}

An EOFC with 2.35-GHz bandwidth and 1-MHz repetition rate is generated to measure the transmission spectrum of a H$^{13}$CN gas cell. The H$^{13}$CN gas cell with a length of 15 cm and a pressure of 25 Torr is in a laboratory temperature of about 297 K. The results shown in Fig.6 (b) are composed of the resolved spectra of 1890 acquisitions. Each acquisition obtained in 1 ms contains 2350 comb lines with a resolution of 1 MHz. Totally 4$\times$10$^6$ spectral points are recorded in the whole spectrum covering a bandwidth of 4.74 THz or 38 nm. The measurement result and a reference database of the $R$ and $P$ branches of H$^{13}$CN are reversely shown in Fig. 6(b) and 6(a), respectively in blue and red. The residual errors between the database and the result are shown in Fig. 6(c). The standard deviation is calculated to be 0.0069, corresponding to a SNR of 144. 
\begin{figure*}[tb]
\centering
\includegraphics[width=15cm]{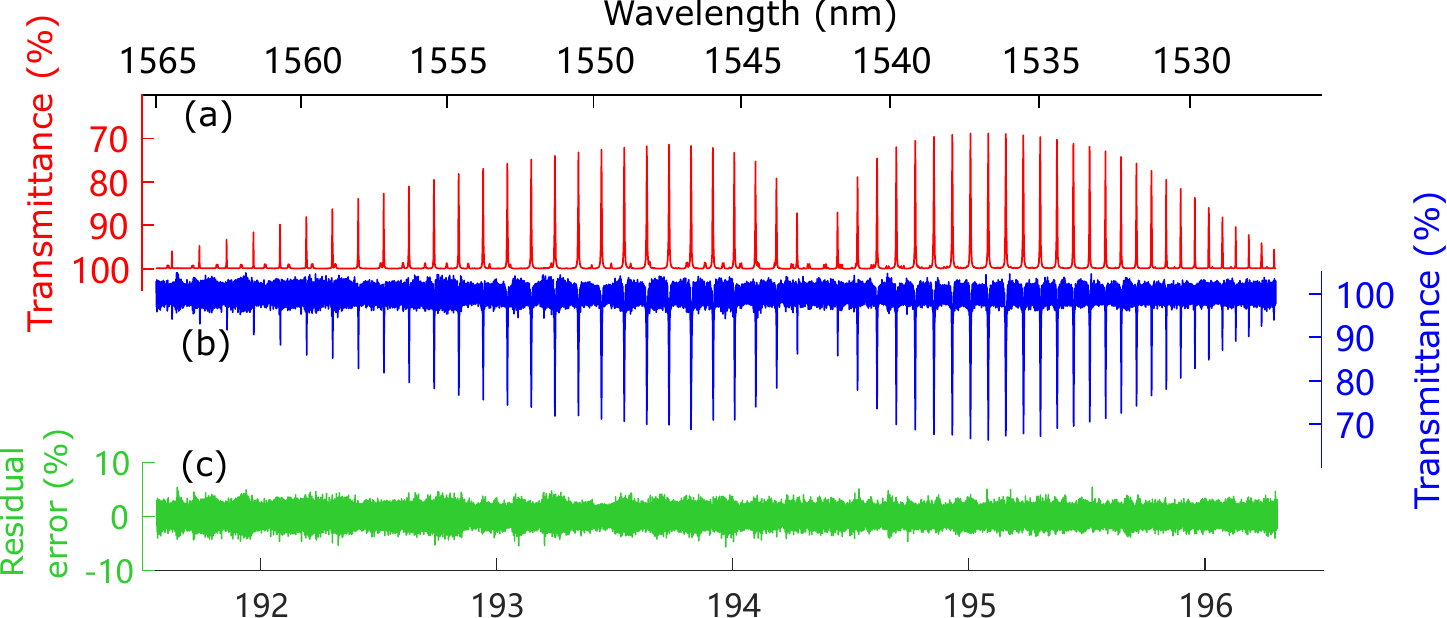}
\caption{\textbf{Spectroscopic measurement of the transmission spectrum of HCN gas cell with 1 MHz resolution.} (a) The reference database and (b) the measurement result of HCN transmittance spectrum in 4.74 THz bandwidth (corresponding to 38 nm) with a resolution of 1 MHz. The result is composed of 1890 acquisitions. Each acquisition contains 2350 resolved comb lines with a line-spacing of 1 MHz in 1 ms recording time. (c) The residual error between (a) and (b). The standard deviation is 0.0069, corresponding to a signal-to-noise ratio of 144.}
\label{fig6}
\end{figure*}

\section{Discussion}
Generally speaking, speckle-based wavemeter and spectrometer suffer from the environmental perturbation, since the speckle patterns generated from multiple interference and scattering light are extremely sensible. Therefore, active stabilization techniques are required for practical measurement, otherwise, the distortion of speckle decreases the precision and accuracy. 
The WGM resonances are also sensible to external temperature and vibration.
All characteristics of resonances, such as linewidths, coupling depths, and resonant wavelengths, are included in WGM barcodes~\cite{2021liao}, which resists mild nonlinear response of moderate external perturbation.
Therefore, the frequency calibration, conducted by measuring the coupling of the ultra-stable laser and the measurement laser, enables our measurement without active stabilization of the resonator. 
The WGM speckle-based scheme proposed in this paper can also be independently used as a wavemeter, which can determine absolute frequency with a resolution of 0.8 fm and a bandwidth of 80 nm.

In speckle-based spectrometer or wavemeter, optical path length is usually longer than other conventional spectrum analyzers with the same device size since multiple interference or scattering, which facilitates miniaturization and high resolution. The inverse relation between FWHM of the cross-correlation peak and Q-factor in WGM speckle-based wavemeter also demonstrates that higher resolution can be achieved with a longer optical path length (details are in Supplement 1). Compared with other structures, WGM micro-resonator is able to realize longer optical path length with same footprint due to high Q-factor. Therefore, we believe that WGM micro-resonator is a potential choice to achieve stability, miniaturization, and high resolution for speckle-based spectrometer or wavemeter.

Here, we also propose and experimentally realize a novel spectroscopy method that introduces ultra-fine EOFC to exploit the high resolution and broad bandwidth of speckle-based wavemeter for spectroscopic measurement. Ultra-fine EOFC precisely samples the spectrum, whose centre frequency is absolutely determined at the same time. This scheme breaks through the bandwidth limitation of EOFC in high-resolution spectroscopy by stitching a sequence of spectral acquisitions centred at different frequencies. Such combination of EOFC and speckle-based wavemeter fully utilizes each characteristics and advantages to be a prospective way towards ultra-high resolution spectroscopic measurement over an much broader bandwidth for precise investigation in bio-chemical sensing and physics.


\subsection* {Acknowledgements}
We thank Prof. Yuejian Song and Dr. Jun Li from Nanjing University for the fabrication of the resonator. This work is financially supported by National Natural Science Foundation of China (NSFC) under grant 62275151. The authors declare no conflicts of interest.

\subsection* {Data, Materials, and Code Availability} 
The data that support the findings of this study are available from the corresponding author upon reasonable request.


\bibliographystyle{ieeetr}   




\listoffigures

\end{spacing}
\end{document}